\documentclass[pra,twocolumn]{revtex4}
\usepackage{bbm}
\usepackage{natbib}
\usepackage{graphics}
\usepackage{amsmath}
\usepackage{mathrsfs}
\usepackage{theorem}
\usepackage{float}
\usepackage{hyperref}
\usepackage{rotating}
\usepackage{amssymb}
\usepackage[english]{babel}
\usepackage{color,times}
\usepackage{CJK}
\usepackage{multirow}
\usepackage{lscape}

\newcommand{\ket}[1]{|#1\rangle}

\newcommand{\be}{\begin{eqnarray}}
\newcommand{\ee}{\end{eqnarray}}

\begin{document}
\begin{CJK*}{GB}{gbsn}
\title{Bound state and Localization of excitation in many-body open systems}
\author{H. T. Cui(´Þº£ÌÎ)$^ {a,b}$}
\email{cuiht@aynu.edu.cn}
\author{H. Z. Shen (ÉòºêÖ¾)$^ {a}$}
\author{S. C. Hou (ºîÉÙ³Ì)$^{a}$}
\author{X. X. Yi(ÒÂѧϲ)$^{a}$}
\email{yixx@nenu.edu.cn}
\affiliation{$^a$Center for  Quantum Sciences, Northeast Normal University, Changchun 130024, China}
\affiliation{$^b$School of Physics and Electric Engineering, Anyang Normal University, Anyang 455000, China}
\date{\today}

\begin{abstract}
Bound state and time evolution  for single excitation in one dimensional XXZ spin chain within non-Markovian reservoir are studied exactly. As for bound state, a common feature is the localization of single excitation, which means the spontaneous emission of excitation  into reservoir is prohibited. Exceptionally  the pseudo-bound  state can always be found, for which the single excitation has a finite probability emitted into reservoir. We argue that under limit $N\rightarrow \infty$ the pseudo-bound bound state characterizes  an equilibrium between the localization in spin chain and spontaneous emission into reservoir. In addition, a critical energy scale for bound states is also identified, below which only one bound state exists and it also is pseudo-bound state. The effect of quasirandom disorder is also discussed. It is found in this case that the single excitation is more inclined  to locate at some spin sites. Thus a many-body-localization like behavior can be found. In order to display the effect of bound state and disorder on the preservation of quantum information, the time evolution of single excitation in spin chain studied exactly by numerically solving the evolution equation. A striking observation is that the excitation can be stayed at its initial location with a probability more than 0.9 when the bound state and disorder coexist. However if any one of the two issues  is absent, the information of initial state can be erased completely or becomes mixed. Our finding  shows  that the combination of bound state and disorder can provide an ideal mechanism for quantum memory.
\end{abstract}

\maketitle
\end{CJK*}

\section{Introduction}
Bound state in open systmes was first defined and studied in photonic material, in which the level of the embedded atom  as an impurity,  was dressed by radiation field \cite{john}. Physically the energy of atom-photon bound state lies in the photonic band gap, and thus the excited photon is exponentially localized in the vicinity of atom. The existence of atom-photon bound state is an universal feature of photonic material, independent of the fine structure of atom. Recently this issue is reconsidered on the ground point of open system. In this case the bound state can still be defined when the spectral density vanishing \cite{apb, twoqubit} or a finite band occurring \cite{zheng,  twoqubit2}.  Similar to the case in photonic materials, the bound state is responsible to the vanishing spontaneous emission, and thus can be used  to protect  the system against decoherence. Experimentally the bound state has been verified in photonic crystals \cite{bsexp}, in which both inhibited and enhanced decay rates can be controlled by the crystals lattice parameter.

Recently the bound state in open two-qubit systems has received extensive interest, focusing on the preservation of quantum information \cite{twoqubit, twoqubit2}. It is known now that the existence of bound state can be used to protect  entanglement against decoherence. Furthermore in topological  two-band systems, the bound state  can also  be found  when the system is coupled with environment, which is responsible to the robustness of  Hall conductance \cite{shen}. In addition the bound state for  cold atoms in optical lattice has been studied \cite{btcold}, which provides an alternative way to controlling the atomic state. A generalization of bound state into multiple levels is also discussed in Ref. \cite{shi}

However a general understanding of bound state in open systems seems absent since the previous works focused mainly on small systems. Consequently  it  is an interesting issue whether there exists bound state in many-body case, and  what the difference is. For this purpose the bound state in one-dimensional XXZ spin chain within non-Markovian environment is discussed in this article. For exactness, only single excitation in spin chain is discussed. Although of simplicity, single-excitation in spin chain has extensive application in quantum information transfer \cite{kay}. Moreover it is shown recently that universal quantum computation can be realized in sing-excitation subspace \cite{geller}. Recently a mechanism for quantum spin lenses has been proposed, in which the spin excitation can be focused on a definite site in one-dimensional chain for storage of quantum information \cite{glaetzle}. With respect of these applications our consideration  has extensive  interest.

Our study shows that $N$ bound states at most can be constructed , in which $N$ is the spin number in  one-dimensional XXZ spin chain. Except of a special  case, the bound state generally  displays  the  localization of single excitation in spin chain, which inhibits the spontaneous emission of spin excitation into environment. However an exceptional case can be identified, in which the excitation is inevitably absorbed  with a finite probability by environment. Furthermore it is found  for $N \rightarrow \infty$ that the probability tends to be 0.5, which corresponds to an equilibrium between the system and environment.  Thus we argue that this state is not a true bound state and is named as pseudo-bound state in this article.

With respect of the recent interest in many-body localization, the effect of disorder in XXZ model is also studied. The appearance of disorder enhances greatly the ability of bound state to protect  excitation against spontaneous emission.  This feature is   attributed to strong localization in bound state, for which the distribution of excitation in spin chain becomes more pronounced in some spin sites. We argue that this phenomena is guaranteed by the combination of bound state and disorder. Thus the environment induced many-body localization can be identified. This finding  is counter-intuitional since  incorporation of environment induces effectively additional long-range hopping, which is  believed to  destroys localization \cite{hopping}.  Moreover  it suggested that the bound state in open many-body systems would  be localized, and furthermore the localization can be strengthened  by  disorder in system. Consequently the combination of bound state and disorder  provided an ideal platform for preservation of information in quantum systems.

The discussion is divided into five sections. In Sec.II the model and definition of bound state are introduced. In Sec. IIIA and IIIB, bound states is evaluated explicitly up to $N=12$ without disorder. The effect of disorder is studied solely in Sec. IIIC. Except of some special cases, the exact calculation becomes unstable for larger $N$. Although of this deficiency, some universal features  in bound state can be found. In order to highlight the crucial role of disorder and bound state in preserving quantum information, the time evolution of single excitation in spin chain is discussed in Sec.IV. Finally conclusion and  furtherm discussion are presented in Sec. V.

\section{Model}

Consider the one-dimensional XXZ model coupled to zero temperature reservoir, of which the Hamiltonian is
\begin{align}\label{h}
H=& \frac{J}{2}\sum_{i=1}\left(S_i^+ S_{i+1}^- + S_i^- S_{i+1}^+ \right) + \sum_{i=1}^{N} h_i S_i^z + U \sum_{i=1} S_i^z S_{i+1}^z \nonumber\\
&+ \sum_{k}\omega_k a^{\dagger}_k a_k + \sum_{k, i}\left(g_k a_k S_i^{+} + g_k^* a_k^{\dagger} S_i^- \right),
\end{align}
with $\hbar \equiv 1$ and the spin number $N$. $S_i^{\pm}$ and $a_k^{(\dagger)}$ are the creation/annihilation operators for  spin $1/2$ and the $k$-th model with frequency $\omega_k$ in the reservoir. $J$ is the tunnelling strength between nearest-neighbor sites. $U$ characterizes the Ising interaction. The on-site field $h_i$ can be homogeneous or randomly distributed, which has  distinct effect on bound state as shown in the following discussion. Periodic and open boundary condition in spin chain are considered respectively. It should be noted that since the existence of Ising interaction the complete spectrum of one-dimensional XXZ model cannot be determined analytically, and one has to rely on the numerical method.

The bound state in open systems is defined as  the energy eigenstate of the total Hamiltonian, which satisfies the equation,
\be\label{bs}
H\ket{\psi_E}= E \ket{\psi_E}.
\ee
Generally the bound state characterizes a bipartite entanglement between system and its environment. Thus when the measure of entanglement is vanishing, the system is  decoupled from environment since the bound state is completely separable in this case. In this sense, decoherence free subspace is a special case of the bound state \cite{dfs}.  In the general case that entanglement of bound state is finite, the coherence in system could be protected partially against decoherence.

With respect for the exactness, only single-excitation situation is considered for determination of $\ket{\psi_E}$ in this article. For more excitation Eq. \eqref{bs} becomes very complex and an exact evaluation is impossible \cite{shi}. Thus $\ket{\psi_E}$ can be written generally as
\be\label{psie}
\ket{\psi_E}&=& \left(\sum_{i=1}^N \alpha_i \ket{\uparrow}_i\ket{\downarrow}^{\otimes (N-1)} \right)\otimes \ket{0}^{\otimes M} + \nonumber \\ &&\ket{\downarrow}^{\otimes N} \otimes \left(\sum_{k=1}^{M} \beta_k \ket{1}_k\ket{0}^{\otimes (M-1)} \right),
\ee
in which $\ket{\uparrow(\downarrow)}_i$ is the eigenstate of $S^z_i$ with the eigenvalue $\pm 1/2$, $\ket{0}_k$ is the vacuum state of $a_k$ and $\ket{1}_k=a_k^{\dagger}\ket{0}_k$, $M$ the number of mode in reservoir. Then substituting the expression of $\ket{\psi_E}$ into Eq. \eqref{bs}, one obtains
\addtocounter{equation}{1}
\begin{align}\label{bsea}
\frac{J}{2}\left(\alpha_{i+1} + \alpha_{i-1}\right) + \left(h_i - U\right)\alpha_i + \sum_{k=1}^{N_k} g_k \beta_k = E \alpha_i  \tag{\theequation a} \\
\omega_k \beta_k + g_k^*\sum_{i=1}^N \alpha_i= E \beta_k \tag{\theequation b}\label{bseb}
\end{align}
in which the resulted constants has been incorporated into $E$.  By Eq. \eqref{bseb},
\be
\beta_k=\frac{ g_k^*}{E-\omega_k}\sum_{i=1}^N \alpha_i.
\ee
Substitute the relation above into Eq.\eqref{bsea}, one obtains
\be
\frac{J}{2}\left(\alpha_{i+1} + \alpha_{i-1}\right) + \left(h_i - U\right)\alpha_i + \left(\sum_{k=1}^{N_k} \frac{ \left|g_k\right|^2}{E-\omega_k}\right) \sum_{i=1}^N \alpha_i= E \alpha_i.\nonumber
\ee
With respect of continuum spectrum in reservoir,
\be\label{int}
\sum_{k=1}^{N_k} \frac{ \left|g_k\right|^2}{E-\omega_k}= \int_0^{\infty} \frac{J(\omega)}{E-\omega} \text{d}\omega,
\ee
in which the spectral density $J(\omega)= \sum_{k=1}^{N_k}  \left|g_k\right|^2 \delta\left(\omega-\omega_k\right)$. Then
\be\label{boundeqn}{\small
\tfrac{J}{2}\left(\alpha_{i+1} + \alpha_{i-1}\right) + \left(h_i-U\right)\alpha_i + \sum_{i=1}^N \alpha_i\int_0^{\infty} \text{d}\omega\tfrac{J(\omega)}{E-\omega} = E \alpha_i}
\ee

It is obvious that the integral in Eq. \eqref{int} is divergent when $E>0$. Thus bound state can exist only when $E<0$ (For complex $E$, the imaginary of $E$ means dissipation and thus bound state cannot exist in this case ). For concreteness the following spectrum function of environment is considered in this article
\be
J(\omega)= \eta \omega \left(\frac{\omega}{\omega_c}\right)^{s-1}e^{-\omega/\omega_c},
\ee
in which $\eta$ characterizes coupling strength between the system and environment, and $\omega_c$ is the cutoff of the environment spectrum. The environment can be classified by  $s$ as sub-Ohmic ($s<1$), Ohmic ($s=1$) and super-Ohmic ($s>1$), which characterizes different relaxation \cite{leggett}. It has known that  a critical coupling constant $\eta_c(s)$ can be defined for single spin ($N=1$) by setting $E=0$ in Eq. \eqref{boundeqn}  \cite{twoqubit2}. Thus bound state can appear only when $\eta>\eta_c$. In this meaning  two distinct phases can be defined in open system, dissipative phase and non-dissipative  phase, which can be transformed each other by a non-equilibrium phase transition \cite{twoqubit2}. However the situation  becomes different  when $N\geq 2$. Obviously now $\eta_c$ is  dependent on the coefficient  $\alpha_i$ as shown in Eq.\eqref{boundeqn}, which suggested that there would exist more than one $\eta_c$. Actually one can find $N$ solutions at most by solving Eq. \eqref{boundeqn} with properly chosen parameters, as shown in the following. However not all solution correspond  genuine bound states. We first point out in this article that there always exists a \emph{pseud-bound-state}, which satisfies Eq. \eqref{boundeqn} but display a relatively high probability  for  single excitation spin being absorbed by environment.

\section{Bound state}

The sufficient and necessary condition for nontrivial  solution to $\alpha_i$  in  linear system of equation Eq. \eqref{boundeqn} is that the determinant of  coefficient matrix is zero. Thus one can obtains an equation for  variable  $E$ with the maximal power of $N$. Generally it is difficult for $E$  to find an analytical expression, and thus  one  has to rely on the numerical method. However because of the limit of computer performance, our numerical evaluation of bound state is restricted to $N\leq 12$ ( the numerical evaluation become unstable  for larger $N$).

The crucial feature of bound state  is the vanishing spontaneous emission of excitation. In order to display this point, two distributions, defined as
\be\label{bsxishu}
c_i&=&\left|\alpha_i\right|^2; \nonumber \\
d&=&\sum_k \left|\beta_k\right|^2 = \left|\sum_{i=1}^N\alpha_i\right|^2\int_0^{\infty} \frac{J(\omega)}{(E-\omega)^2} \text{d}\omega,
\ee
are calculated exactly, in which $c_i$ characterizes the probability of excitation located  at the $i$-th spin site, and  $d$ is the probability of excitation absorbed by reservoir. In addition $d + \sum_{i=1}^N c_i =1$ is imposed.  For brevity, the following discussion focuses only on the case of Ohmic environment ($s=1$). As for sub- and super-Ohmic cases, our evaluations shows no essential difference from $s=1$.

\begin{figure*}
\center
\includegraphics[width=13cm]{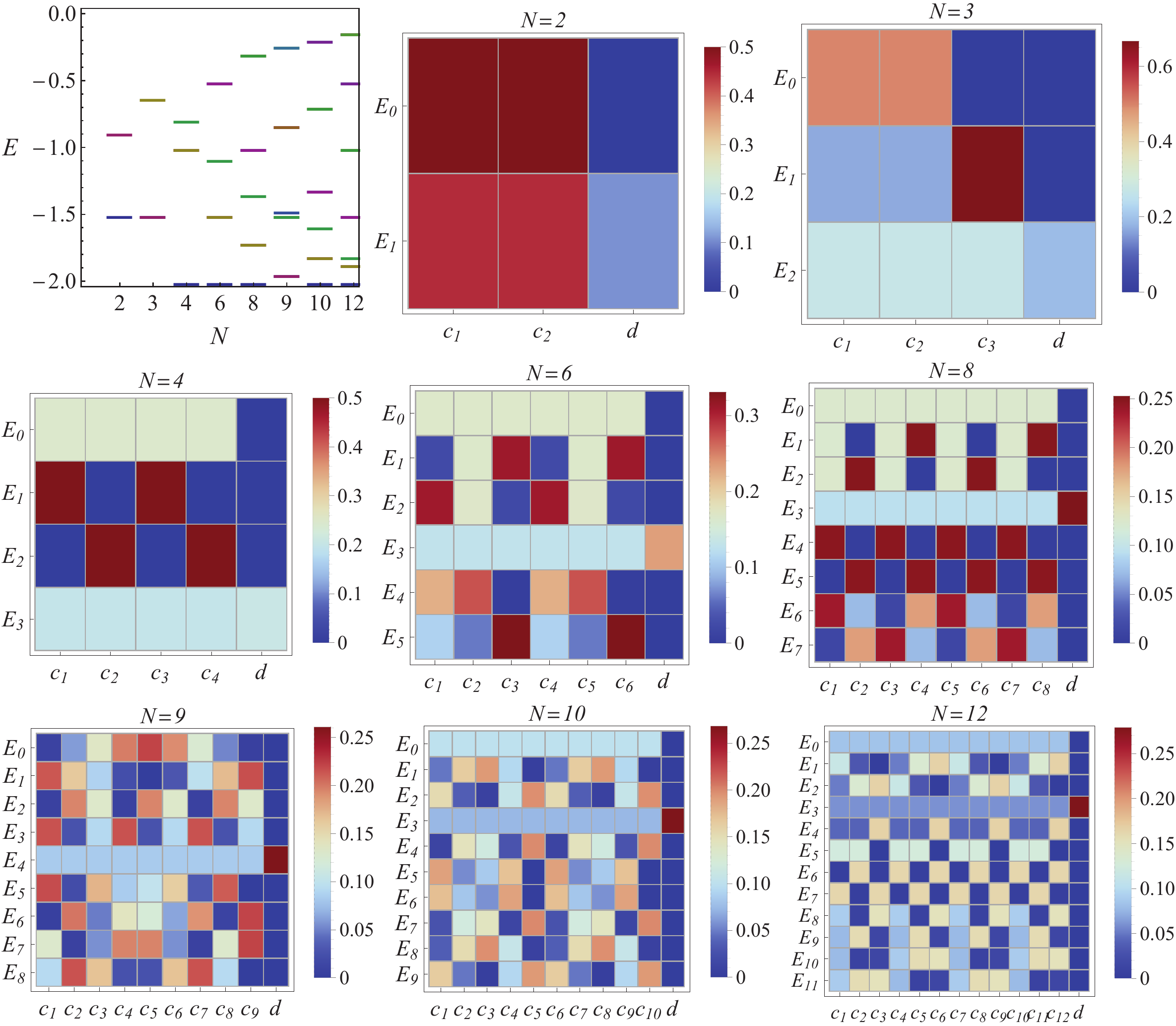}
\caption{(Color online) A plot of  eigenenergy $E$ and density plots of normalized distribution $c_i$ and $d$ in  bound states for $N=2$ to $N=12$  with periodic boundary condition. $h_0 - U=-1$, $s=1$, $\eta=0.1, \omega_c=3 $ in unit of $J$ have been  chosen for this plot. The eigenenergy $E_n (n=0, 1, 2,\cdots)$ is arranged by increased order.  }
\label{fig:ps1}
\end{figure*}

\subsection{Periodic boundary condition}

With respect of periodic boundary condition in spin chain,  $S_{i+N}= S_i$ and $h_i=h_0$ are imposed. By proper choice  of $h_0-U$, one can find $N$ solution of $E$ at most for definite $N$. In Fig. \ref{fig:ps1},  $c_i$ and $d$ are plotted  for different $N$ when $s=1$ respectively. A common feature is   the  periodic variation of $c_i$ by spin site  when  $N \geq 4$,  which can be attributed to the spin-site translational invariance in Eq.\eqref{h}. Furthermore it is not difficult to note the periodicity in  $c_i$ is decided by decomposition $N=m\times n (m, n>0)$.

An interesting observation is  the double degeneracy in bound energy level $E_n (n\geq 2)$ when $N\geq 3$. In addition an even-odd effect can  be found; When $N$ is even,  there exists two non-degenerate levels, the ground state $E_0$ and some excited state. And the other levels are  double degenerate. However when $N$ is odd, there is only one non-degenerate level, the others being double degenerate.

Another  crucial observation is  that there always exist an unique energy level with nonvanishing  $d$, which means that the excitation in spin systems would be emitted into reservoir with a finite probability. While the other bound states  $d=0$ exactly. For instance as shown in Figs. \ref{fig:ps1},  the level $E_2$ for $N=3$ , $E_3$  for $N=4, 6, 8, 10, 12$ and $E_4$ for $N=9$ have $d>0.1$.  Consequently  the appearance of the state would  erase the information of initial state during time evolution. Hence even though satisfying Eq. \eqref{boundeqn}, we argue that it is not a true bound state  since the excitation cannot be localized completely  in spin chain.  Thus it is named as \emph{pseud-bound state} in this article.

\begin{table}
\begin{tabular}[c]{|c |  c c | }
\hline
\multirow{2}*{$N$} & \multicolumn{2}{c|}{$s=1$}  \\
\cline{2-3} & $E$ &  $d$\\
\hline
14 & -2.01932 & 0.289168 \\
15  & -2.12051 & 0.293435 \\
16 & -2.21916 & 0.297389 \\
17 & -2.31547 & 0.301069  \\
18  & -2.40961 & 0.304187 \\
19 & -2.50172 & 0.30773 \\
20 & -2.59159 &  0.310762 \\
21& -2.68041 & 0.313621\\
22& -2.7672 & 0.316325 \\
\hline
\end{tabular}
\caption{\label{table:ps} The bound state with energy $E<-2$ and the corresponding $d$ in periodic boundary condition. The other parameters are same as those in Fig.\ref{fig:ps1}. }
\end{table}

\begin{table}
\begin{tabular}[c]{|c |  c c |}
\hline
\multirow{2}*{$N$}  & \multicolumn{2}{c|}{$s=1$} \\
\cline{2-3} & $E$ &  $d$\\
\hline
100  & -7.33804 & 0.391086 \\
200 & -11.0897 & 0.416524 \\
400  & -16.4958 & 0.436926 \\
800 & -24.2306 & 0.45294  \\
1600  & -35.2459 & 0.46527  \\
3200& -50.9722 & 0.474613 \\
\hline
\end{tabular}
\caption{\label{table:pss} The lowest bound-state energy  $E$ and the corresponding $d$ for larger $N$ in periodic boundary condition. The other parameters are same to those in Fig.\ref{fig:ps1}. }
\end{table}

In addition  the  pseud-bound state displays three critical features.  First the  pseud-bound state always is non-degeneracy. Second our exact calculation displays that the  pseud-bound state actually is a  $N$-qubit $W$ state. Thirdly up to exact calculation up to $N=22$, we find that there is only one bound state when  $U+E <- 1$ or $E<2$ correspondingly with the chosen parameters in Fig.\ref{fig:ps1}. Furthermore it also is a pseud-bound state,  as shown in Table \ref{table:ps}. These facts mean that the ground state for $N\rightarrow \infty$ would be non-degenerate and a  pseud-bound state.

Furthermore $d$ in pseud-bound state increases  with spin number $N$, shown in Table \ref{table:ps}. Thus it is interesting to find the nontrivial upper bound for $d$. With respect that the pseud-bound state is a $N$-qubit $W$ state, it is found  that  $d\rightarrow 0.5 $ when  $N\rightarrow \infty$, as shown in Table.\ref{table:pss} (With respect of  $\alpha_i=1/\sqrt{N}$, Eq. \eqref{boundeqn} is simplified greatly in this case. Thus the energy level can be determined exactly for arbitrary $N$ ). It means that the relaxation and localization of spin excitation become balanced in this case, and thus pseud-bound state is actually an equilibrium state. Moreover  our exact calculation shows that the ground state is the unique pseud-bound state for larger $N$. The appearance of pseud-bound state in excited level for small $N$  shown in Figs. \ref{fig:ps1}, can be attributed  to the finite number effect.

In conclusion  two distinct states can be found in this case: bound state   and pseud-bound state. A critical case can be identified as $U +E =-1$, which separate the true bound state from  pseud-bound state. When $U +E <-1$, there is only one bound state, which also is pseud-bound state. For $U +E >-1$, all energy levels are true bound states with $d=0$ exactly under $N\rightarrow \infty$.  Interestingly   we find  that the bound state, if existing at the critical energy $U + E =-1$ or $E =-2$, also non-degenerate, and  is a W-state like state
\be
\frac{1}{\sqrt{N}}\sum_{i=1}^N \left(-1\right)^i\ket{\underbrace{0\cdots 0}_{i-1} 1 \underbrace{0\cdots 0}_{N-1}}.
\ee
Thus it is a true bound  state for even $N$ since $d = 0$ exactly.


\begin{table}
\begin{tabular}[c]{|c |  c c | }
\hline
\multirow{2}*{$N$}  & \multicolumn{2}{c|}{$s=1$}  \\
\cline{2-3} & $E$ &  $d$ \\
\hline
13  & -2.01274 & 0.209038 \\
14 & -2.09872 & 0.265101 \\
15  & -2.19033 & 0.277008 \\
16  & -2.28187 & 0.284867 \\
17  & -2.37255 & 0.290923 \\
18  & -2.46206 & 0.295991 \\
19  & -2.55026 & 0.300411 \\
20 & -2.63714 & 0.304362 \\
21  & -2.72269 & 0.30795 \\
22 & -2.80692 & 0.311247\\
\hline
\end{tabular}
\caption{\label{table:os} The bound state with energy $E<-2$ and the corresponding $d$ in open boundary condition. The other parameters are same to that in Fig.\ref{fig:os1} in order to find $d$. }
\end{table}

\begin{figure*}
\center
\includegraphics[width=13cm]{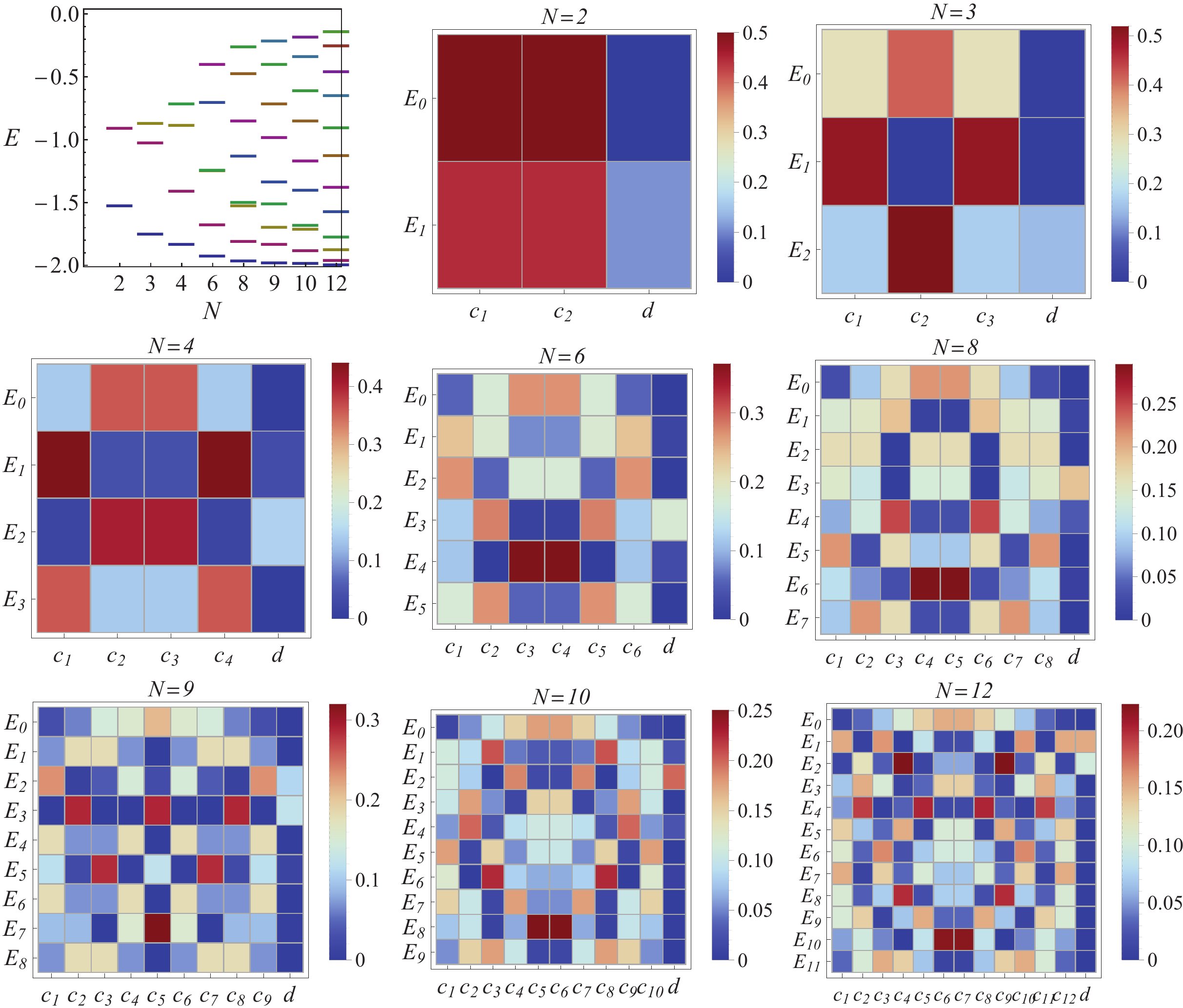}
\caption{(Color online) Plots for energy level $E$, $c_i$ and $d$ when $s=1$ for open boundary condition. The other parameters are same to those in Fig.\ref{fig:ps1}. }
\label{fig:os1}
\end{figure*}

\subsection{Open boundary condition}

\begin{figure}[t]
\center
\includegraphics[width=7cm]{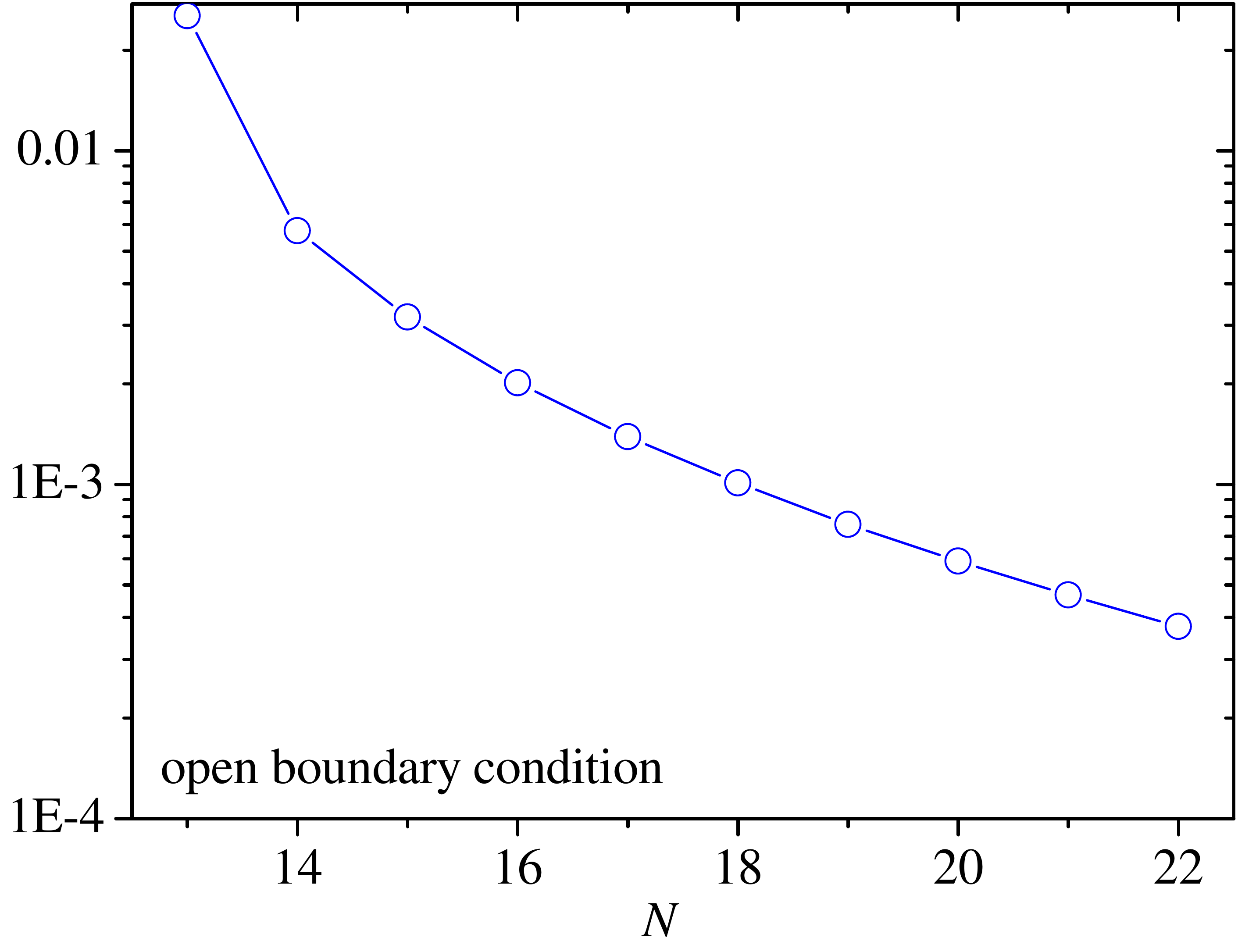}
\caption{ The isotropy of $c_i$, measured by $\sum_{i=1}^N\left(\left|\alpha_i \right|-1/\sqrt{N}\right)^2/N$. The parameters are chosen same as that in Fig.\ref{fig:ps1}.   }
\label{fig:oss}
\end{figure}

For open boundary condition, translational invariance in spin chain is broken. In Figs.\ref{fig:os1}, the energy level and the density plotting for $c_i$ and $d$ with different $N$ are presented. A direct feature  is  broken of degeneracy in bound state, which means  that the double degeneracy  is protected by translational invariance in spin chain. Although the periodicity in  $c_i$ disappears,  a  symmetry to the center of spin site is developed instead, as shown in Fig.\ref{fig:os1}.

In contrast to periodic boundary, the difference between  pseud-bound state and true bound state  become ambiguous since $d$ could  be very small but not vanishing exactly. However the critical energy $U+E_n = -1$ or $E_n=-2$ can still be identified, below which there is only  one bound state, which is also pseud-bound state. In Table. \ref{table:os} the ground-state energy level with $E<-2$ is listed up to $N=22$, in which $d$  increases with the spin number $N$. Moreover the exact evaluation show that the distribution $c_i$ tends to be homogenous as for all spin sites with increment of $N$, shown in Fig. \ref{fig:oss}. It implied that the ground state would also an equilibrium state when $N\rightarrow \infty$.

\subsection{Effect of Disorder in XXZ model}

\begin{figure*}
\includegraphics[width=18cm]{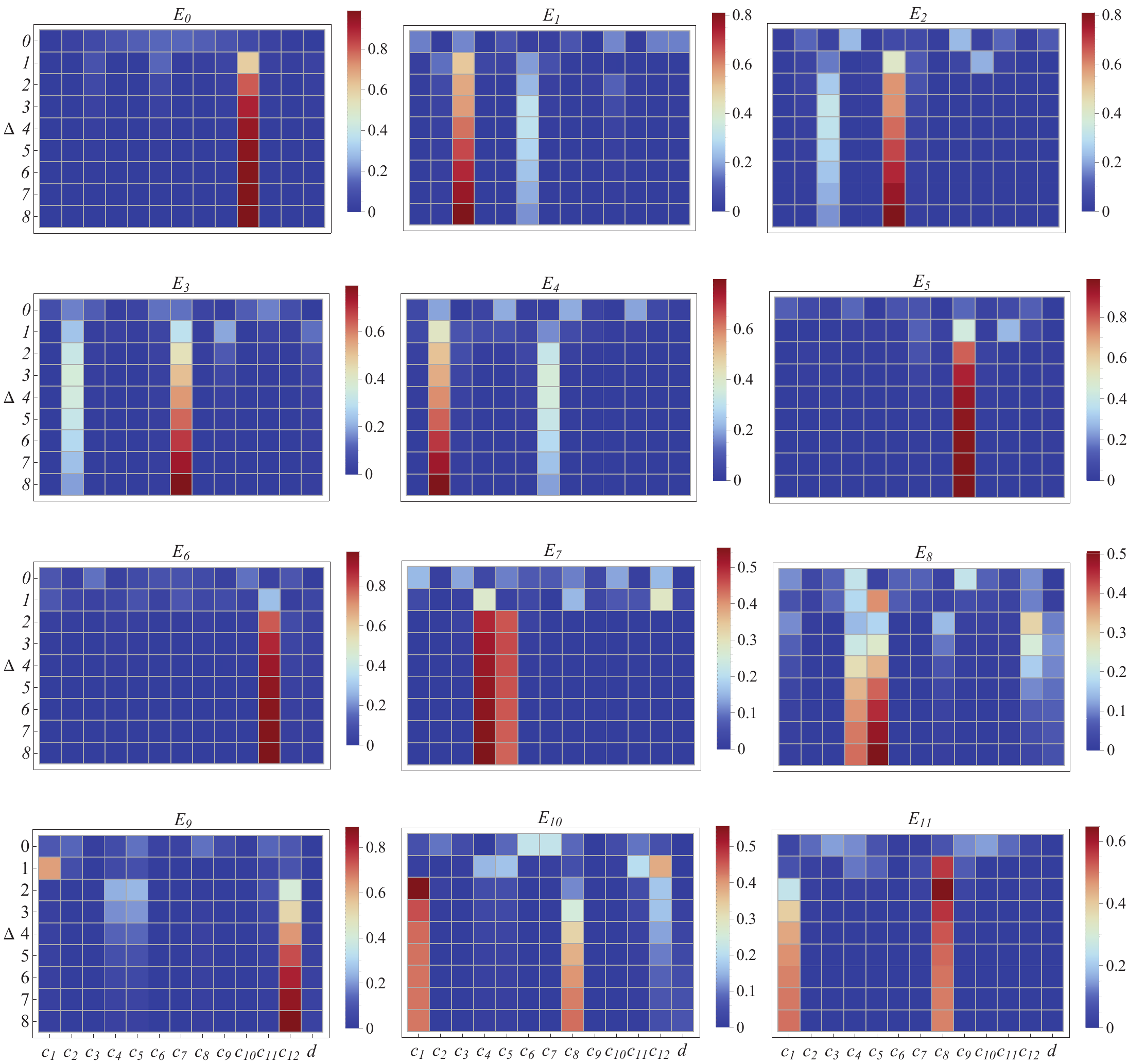}
\caption{(Color online)  The density plots  of  normalized distribution  $c_i$ and $d$ of bound-state energy levels  for $N=12$ with open boundary condition when $s= 1$. For $\Delta=1$ to $8$, $U=1, 1.5, 2.5, 3.2, 4.2, 5.2, 6.2, 7.2, 8.2$ are chosen respectively in order to find all bound states. The other parameters are  $\eta=0.1, \omega_c=3 $ in unit of $J$.  }
\label{fig:n12}
\end{figure*}

It is an interesting issue whether the disorder in spin system would has effect on bound state. This consideration comes from the recent interest in many-body localization (MBL) \cite{mbl}, which  characterizes a nonthermalized behavior in the statistical mechanics of isolated quantum systems. A general feature for MBL is the disorder-induced  non-equilibrium properties, such as  localization of electronic wave function \cite{al} or the deviation of statistical  properties from the expectation of thermodynamical equilibrium\cite{huse07}. As shown in the previous discussion,  true bound state  displays an obvious non-equilibrium since  spontaneous emission of spin excitation into reservoir is inhibited  and excitation is localized in spin chain. From this point the spin system  cannot be thermalized by its reservoir when bound state occurring. Consequently it is an interesting issue whether  the disorder in spin chain could enhance the localization of excitation.

With respect of recent experiments on MBL \cite{mblexp, mblopen1}, a quasirandom disorder is introduced in spin chain with open boundary,
\be\label{disorder}
h_i= \Delta \cos \left(2\pi\beta i + \phi\right),
\ee
in which $\beta=\frac{532}{738}\approx 0.721$, $\phi=1/0.6188333$ and $\Delta$ characterize the strength of disorder. Although $h_i$ does not denote a genuine disorder, the intrinsic effect of disorder can be simulated and demonstrated in this system \cite{mblexp, mblopen1}.   $c_i$ and $d$ of   bound states  for $N=12$ are plotted in Fig. \ref{fig:n12}. It is obvious that $c_i$ becomes so pronounced at some spin sites  with the increment of disorder,  which means that the  spin excitation cannot be transported freely  in spin chain.  Furthermore since the system is open, localization of excitation in this case is stable against decoherence. This picture is very different from the intuition that the reservoir would always destroy localization, as shown for MBL in \cite{mblopen1, mblopen2, steadystate}. We argue that the existence of bound state is responsible for the stability of localization.

As for $d$, the disorder shows finite effect. For instance when $N=12$, $d$ displays obvious variance with $\Delta$ as shown for bound states $E_8$   and $E_{10}$  in Fig. \ref{fig:n12}. So in this case the boundary between true and pseudo bound state is still ambiguous. Finally we stress that our exact examination shows similar  behaviors for $N<12$, which does not present here for brevity.

Importantly  one should note that under single excitation case, Eq. \eqref{boundeqn} actually characterizes a free particle system within reservoir  since the interaction $U$ only contributes to the diagonal elements. Thus it suggested that Anderson localization (AL) would happen in this situation \cite{al}. An essential difference between AL and MBL is that the localization of  wavefunction around some lattice site can be observed in AL, as shown in experiments with ultracold atoms \cite{alexp}. Whereas as shown in Fig.\ref{fig:n12} $c_i$  can be dominated  at more than one spin site in this case, and thus the spin chain would be more entangled. Then a MBL-like picture is observed although there is no effective interparticle coupling.

\section{Time  Evolution}

Localized system is of interest as possible quantum memory since some local details of the system's initial state can be preserved. Thus it is   very interesting how the system evolves in time  when bound state and disorder occur.  By Schr\"{o}dinger equation, one can obtain
\be\label{evolution}
\mathbbm{i}\frac{\partial }{\partial t}\alpha_i(t)&=& \frac{J}{2}\left[\alpha_{i+1}(t) + \alpha_{i-1}(t)\right] + \left(h_i - U\right)\alpha_i(t) \nonumber\\
&&- \mathbbm{i} \sum_i \int_0^t \text{d}\alpha_i(\tau)f(t-\tau),
\ee
in which $f(t-\tau)= \frac{\eta}{\omega_c^{s-1}} \frac{\Gamma(s+1)}{\left[\mathbbm{i}(t-\tau) + 1/\omega_c\right]^{s+1}}$ and $\mathbbm{i}$ is imaginary unit. The effect of reservoir comes from the last term, which also characterizes the Non-Markovian memory structure. The following discussion is divided into two part by $\Delta \neq 0$ or $\Delta=0$ since the XXZ model  would  have distinct feature in the two cases.

In general the behavior of $\ket{\psi(t)}$ of spin chain can be understood by the following expressing
\be\label{psit}
\ket{\psi(t)}= \sum_{E_b} \alpha_b\ket{\psi_b} e^{-\mathbbm{i}E_b t} + \int\text{d}E  \alpha(E)e^{-\mathbbm{i}E t } \ket{\psi_E},
\ee
in which the first term denotes the contribution from bound energy level $E_b$, while the integral is the contribution from  the continuum spectrum with $E>0$.  Thus for time evolution, $\ket{\psi(t)}$ would  become oscillating when bound state happens. In contrast a complete dissipation is expected  when there is no bound states. However when disorder appearing, the picture would be deviated from this expectation as shown in the following discussion.

\begin{figure}
\includegraphics[width=8.5cm]{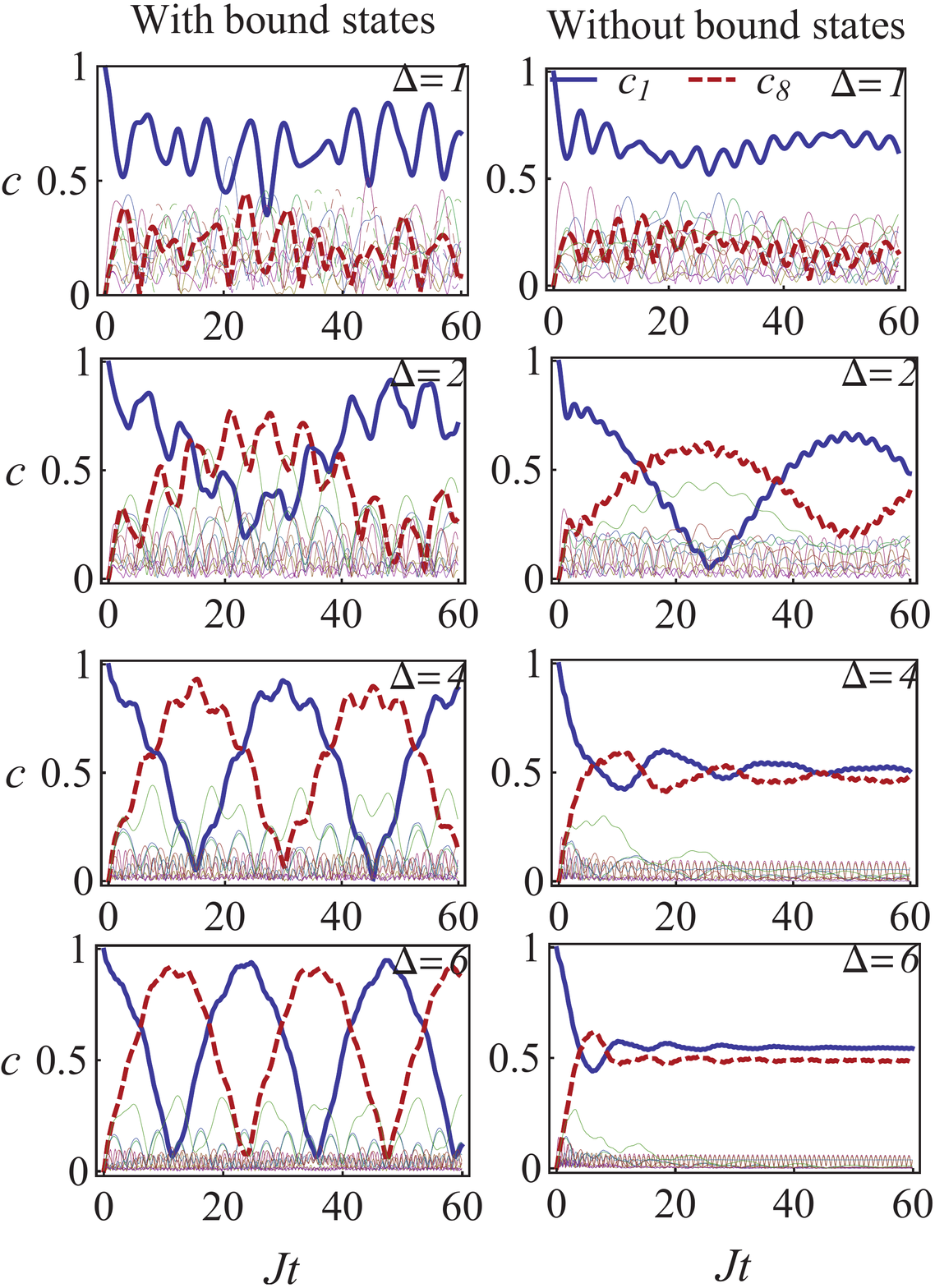}
\caption{(Color online) A plot for evolution of $c_i$  with rescaled time $Jt$ when $N=12$ and $s=1$. This initial state is chosen as the single excitation located on  spin site $i=1$. When all bound states occurring, shown by the left column,  $U=1.5$ for $\Delta=1$,  $U=2.5$ for $\Delta=2$,  $U=4.2$ for $\Delta=4$ and  $U=6.2$ for $\Delta=6$ are chosen; As for the absence of bound state, shown by the right column, $U=-7$ is chosen same for different $\Delta$. $\eta=0.1, \omega_c=3 $ are chosen for this plot. }
\label{fig:on12}
\end{figure}

\begin{figure}
\includegraphics[width=8.5cm]{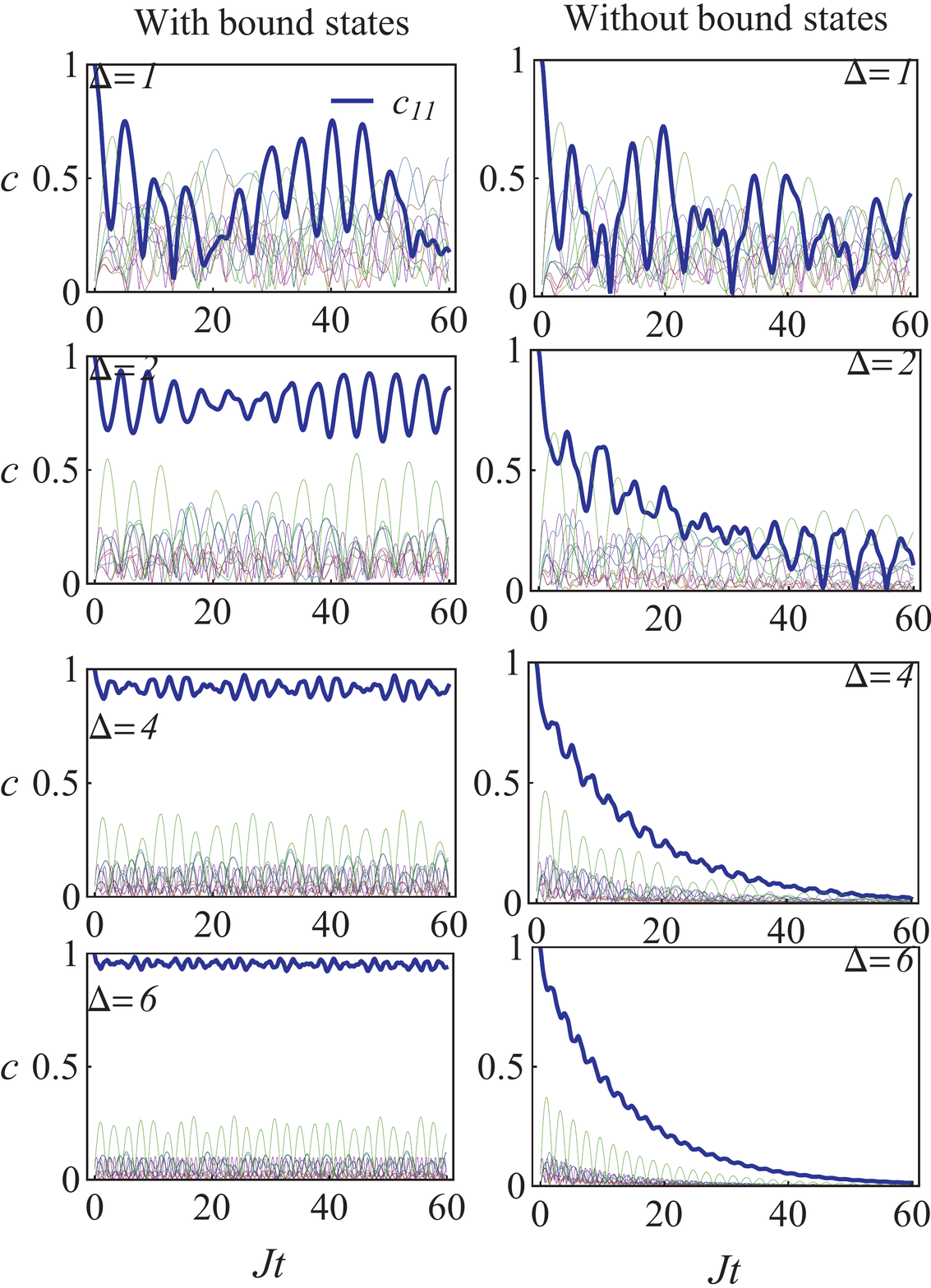}
\caption{(Color online) A plot for evolution of $c_i$  with rescaled time $Jt$ when $N=12$ and $s=1$. Except that  the excitation is initially located at $i=11$, the parameters are same to that in Fig. \ref{fig:on12}. $c_{11}$ is highlighted in this plot.}
\label{fig:r11}
\end{figure}

\begin{figure}
\includegraphics[width=8.5cm]{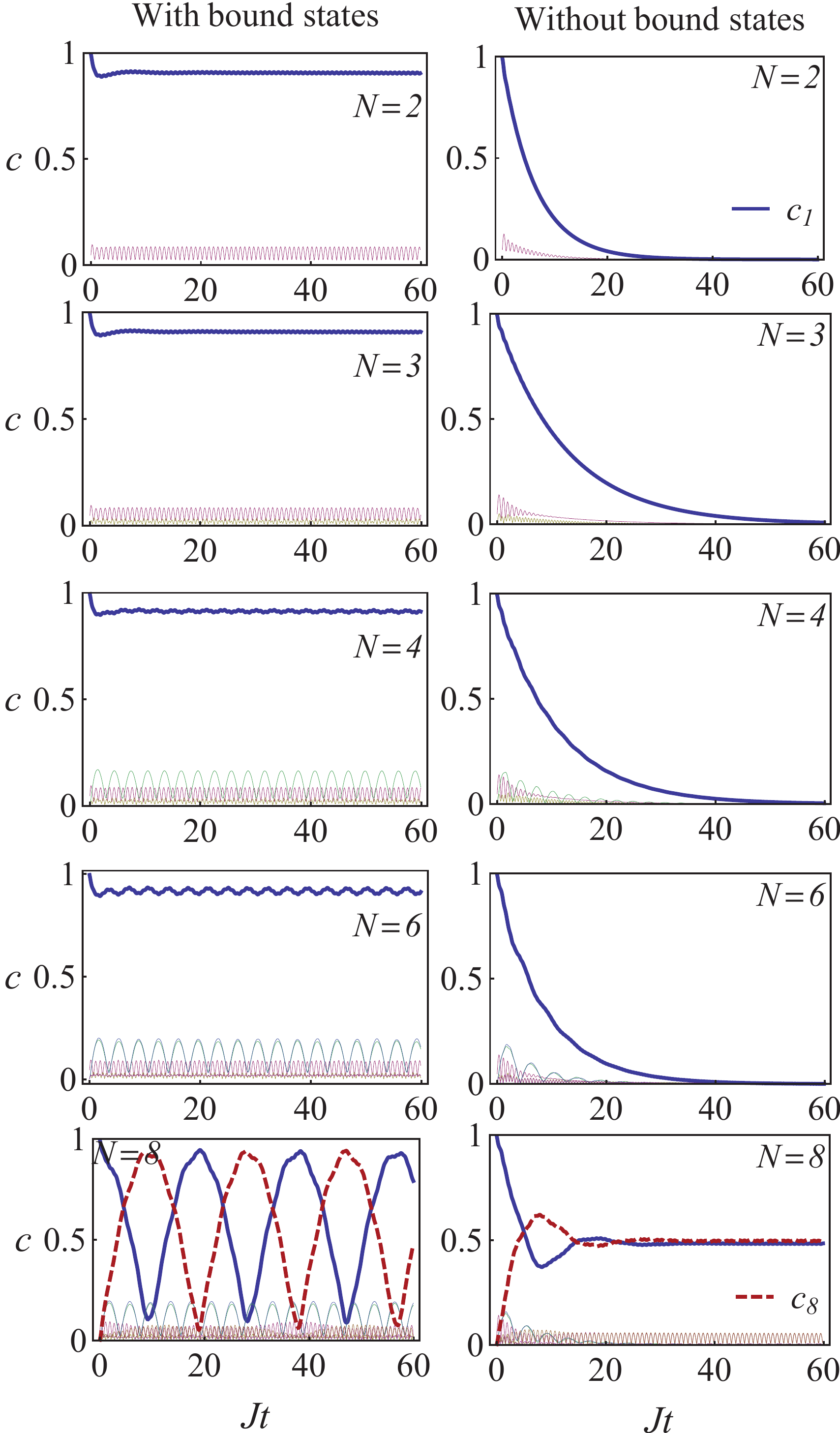}
\caption{(Color online) A plot for evolution of $c_i$  with rescaled time $Jt$ and the spin excitation initially located at $i=1$ for different $N$. For this plot,$\Delta=6$  is chosen and the other parameters are same to that in Fig. \ref{fig:on12}. }
\label{fig:d6}
\end{figure}

\subsection{$\Delta \neq 0$}
By exact numerical evaluation of Eq.\eqref{evolution}, the evolution of $c_i$  for $N=12$ are shown in Fig.\ref{fig:on12} and \ref{fig:r11}  for different $\Delta$ when all bound states appearing or not completely, in which  the spin excitation is located initially at $i=1$ or $i=11$ respectively. For clarity and brevity, the case of $s=1$ is discussed only. Two different pictures can be found.

First when the spin excitation is located initially at $i=1$, a stable  oscillation between $c_1$ and $c_8$ is developed with the increment of $\Delta$ when bound states appearing. This feature can be understood by noting that  $c_1$  and $c_8$  are always pronounced simultaneously as shown for bound states $E_{10}$ and $E_{11}$ in Fig.\ref{fig:n12}. By Eq. \eqref{psit}, the information of initial state can be transferred coherently between the bound states during time evolution. However when there is no bound state,  $c_1$ and $c_8$ become stable simultaneously at the value closed to 0.5. Since the absence of bound state, the feature could be attributed to the effect of disorder.

Recently the open quantum dynamics in localized systems is studied extensively, for which two nonergodic features can be found. In Refs.\cite{steadystate}a localized steady state is disclosed  in open Anderson-localized system, achieved by a proper dissipation. This finding means that the environment could  play a constructive role in localization. Whereas in Refs.\cite{mblopen2}, a stretched exponential decay is discovered in open MBL systems with a general consideration of environment, for which the nonergodic character of system persists for a long time. The underlying mechanism for the exotic dynamics can be attributed to the existence of integrals of motion in localized systems\cite{integrals}, which prevents not only complete thermalization of any given subsystem, but also transport over macroscopic scales. Thus a typical localization length can be defined, which means that the system can be considered as localized when the system scale is much lager than localization length.   Consequently as for the present discussion, the integrals of motion has also important effect on evolution of $c_i$. This point can be illustrated by Fig.\ref{fig:d6}. First consider the case when there is no bound state. It is obvious that  $c_1$ decays rapidly if $N\leq 6$. while a stable behavior for $c_1$ and $c_8$ can be observed if $N\geq 8$, as shown by the bottom-right panel in Fig. \ref{fig:on12} and  \ref{fig:d6}. This feature can be understood by the point that when the localization length is larger than the length of spin chain, the system is inevitably decaying. While with increment of  $N$, the length of spin chain becomes comparable to localization length. So in this case the system displays the  stability  against dissipation. Second the effect of integrals of motion can still be in function  when bound states occurring, as shown by the left column in Fig.\ref{fig:d6}. It is obvious that $c_1$ tends to bo steady at value $\sim 0.9$ when $ N\leq 6$. In contrast a stable oscillation between $c_1$ and $c_8$ is developed when $N=8$. This feature can be understood by that fact that a correlation between distant spin sites can be constructed when the length of spin chain is larger than localization length. Similar phenomenon can be found for $N=12$, as shown in Fig.\ref{fig:on12}.

Second when the spin excitation is located initially at $i=11$, only $c_{11}$ becomes dominant  when bound state and disorder occur together.  As shown by left column in Fig.\ref{fig:r11}, $c_{11}$ settles  rapidly on a value  closed to 0.9 with the increment of $\Delta$ when bound state appearing, while the other $c_i$ tend to vanish instead. This picture can be understood by noting that $c_{11}$ in bound energy level $E_6$ becomes more pronounced with increment of $\Delta$, as shown in Fig.\ref{fig:r11}. Thus by Eq.\eqref{psit} this level have dominant contribution to the first term. And then the overlap between this bound state and initial state is large in this case, while the other $c_i$ tends to be vanishing. In contrast  when there is no bound state, all $c_i$s  tend to be vanishing simultaneously with $Jt$ since the first term in Eq.\eqref{psit} does not occur and the dissipative term becomes dominant, as shown by right column in Fig.\ref{fig:r11}. However $c_1$ decays much slowly with the increment of $\Delta$, which means that disorder alone has finite ability  to preservation of quantum information. This observation means that the disorder itself is not enough for the preservation of quantum information in open system.

In conclusion the two pictures disclose the fact that the system is localized, does not mean that  the time evolution for arbitrary initial state can display nonergodic feature. As shown in this subsection, the nonergodic behavior  happens only for some special initial state. This feature can be understood by the fact that the integrals of motion just determine a finite energy manifold in Hilbert space: only the state in this manifold can behave nonergodically, while the state outside is inevitably thermalized. As for present discussion,  the long time behavior of $\ket{\psi(t)}$  is dependent completely on the location of single excitation at initial time: when the initial location of excitation corresponds to  some bound state, in which only $c_i$ in this location is pronounced with increment of $\Delta$, the information of initial state can be preserved with very high probability when bound states occurring. In contrast it would disappear completely in final when there no bound state. Moreover when there exist another spin site $j$, on which $c_j$ is pronounced simultaneously, the information of initial state can be preserved coherently between $c_i$ and $c_j$ when bound states occurring, while it becomes mixed when there is no bound state. Consequently it seems that the information of integrals of motion could be extracted from the bound state with disorder.

\begin{figure}
\includegraphics[width=8.5cm]{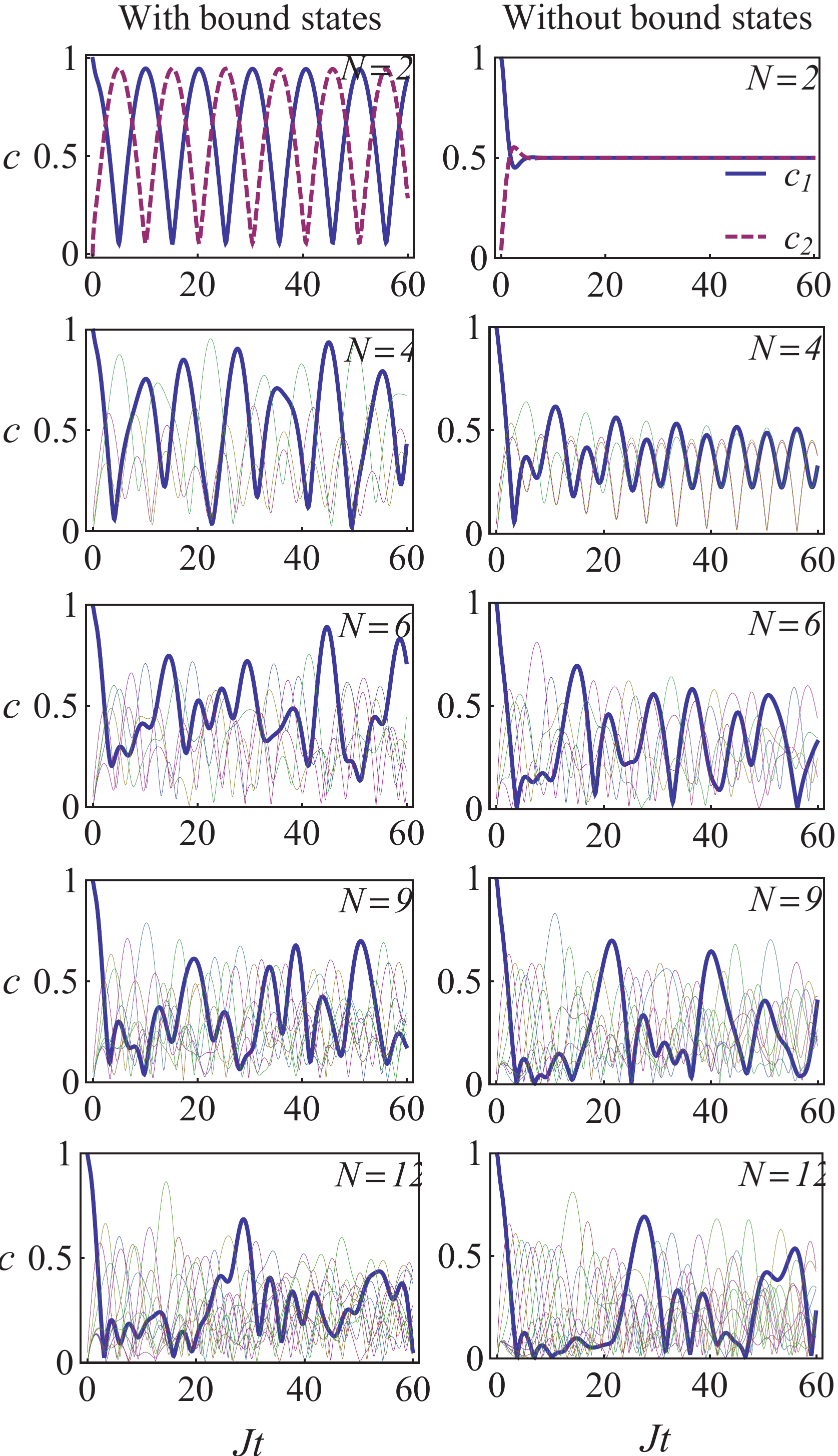}
\caption{(Color online) A plot for evolution of $c_i$  with rescaled time $Jt$ when $\Delta=0$ and $s=1$. This initial state is chosen as the single excitation placed at spin site $i=1$. When all bound states occurring,  $U=1$ ; when there is no bound, $u=-1$ for $N=2$, $u=-1.5$ for $N=4$,$u=-2.5$ for $N=6$,$u=-4$ for $N=9$ and $u=-5.5$ for $N=12$. $\eta=0.1, \omega_c=3 $ are chosen for this plot. }
\label{fig:d0}
\end{figure}

\subsection{$\Delta=0$}

It becomes different when there is no disorder ($\Delta=0$). First consider the case of  bound state happening, For instance when $N=2$, an oscillation happens between $c_1$ and $c_2$,  as shown in Fig. \ref{fig:d0}.  However with the increment of $N$, the behavior of $c_1$ tends to be similar to the other $c_i$.  This picture can be understood by Fig. \ref{fig:os1}, in which the distribution $c_i$  and $d$ are plotted. When $N=2$, there are only two bound states $E_0$ and $E_1$, in which we find $c_1=c_2$. Thus the oscillation is attributed to the coherent superposition of $E_0$ and $E_1$, as shown in Eq. \eqref{psit}.  With the increment of $N$, the difference of $c_i$ in bound states  tends to be decreased. Thus  all bound states tend to have equal contribution to evolution of $\ket{\psi(t)}$ by Eq.\eqref{psit}. Consequently the information of initial state is diluted with time evolution. Thus it could be expected that under $N\rightarrow \infty$ the spin system would become equilibrated eventually. In fact this feature is a  manifestation of eigenstate thermalization hypothesis, which state that all initial states in many-body system would become equilibrium at large time evolution \cite{eth}.

Second when there is no bound state, distinct behavior for evolution of $c_i$ can be found. When $N=2$ shown by the top right panel in Fig. \ref{fig:d0}, $c_1$ and $c_2$ evolve  simultaneously to a steady value closed to  0.5.  However when $N$ is large,  evolution of $c_i$ does not show  intrinsic difference from that bound state appears. In addition our evaluation shows that this feature is insensitive to the initial state and can also exist for periodic boundary condition.

Phenomenally we attribute this phenomenon to the fact that one dimensional XXZ spin chain actually is integrable \cite{integralmodel, rigolPRL}. It means that there exist a number of extensive operators, which is sums of local operators, commuted with the Hamiltonian and with each other. Actually  extensive operator correspond to the conserved quantity or symmetry in system. Thus the time evolution of system is confined to a highly restricted energy manifold \cite{integralmodel, rigolPRL}, which is the underlying mechanism of  resistance to thermalization in integrable system. Consequently the energy manifold can  cover all possible steady state, thus all bound states which is the steady state since it is the eigenstate of the system and environment. This is the reason why the time evolution of system  show negligible difference  when bound state occurring or not for large $N$.



\section{Discussion and Conclusion}

In conclusion the bound state and time evolution for single-excitation in one dimensional XXZ spin chain within reservoir are studied exactly in this article. As for bound state, four crucial observations are found. First the true  bound state is actually a excitation-localized state, for which the single excitation is preserved in spin system and  the spontaneous emission  into reservoir  is prohibited exactly.  Second we  point out firstly the existence of pseudo-bound state, which defined as the state satisfying bound-state equation Eq. \eqref{boundeqn} but having a finite probability of   spontaneous emission for single excitation. Thus in this case the excitation in spin chain would be relaxed probably into reservoir. In addition when periodic boundary condition is imposed in spin chain, pseudo-bound state is unique and exact. The spin chain corresponds to a $N$-qubit $W$ state in this case, that is $\alpha_i=1/\sqrt{N}$ in Eq. \eqref{psie}. Moreover for large $N$ pseudo-bound state corresponds to the minimal energy level and $d$ tends to 0.5, as shown in Table \ref{table:pss}. As for open boundary in spin chain, the boundary between  true and pseudo bound  becomes ambiguous.  Thirdly a critical energy  $U + E =-1$ can be identified, below which there is only one bound state. For periodic boundary, the bound state is pseudo-bound state exactly. As for open boundary, we find that the values all $c_i$ in bound state tend to be  isotropic  as shown in Fig.\ref{fig:oss}, and the value of $d$ is increased with the increment of spin number $N$, shown in Table \ref{table:os}. Fourthly when the quasirandom disorder Eq.\eqref{disorder} occurs,  $c_i$ in bound state becomes more pronounce at some spin sites.  Thus an environment induced many-body localization-like behavior can be observed. This feature implies that single excitation can  be localized in some spin sites even though the existence of environment. Thus the spin chain in this case can be used as quantum memory.

In order to display the potential application in quantum memory, time evolution of single excitation is evaluated exactly by numerics.  A crucial observation is that the information of initial state can be faithfully preserved up to 90\% maximally only if  the bound state and disorder occur together. Moreover the long time behavior of spin chain  is determined completely by the structure of bound state. When there is on bound state, the information of initial state can be erased completely or become mixed, dependent on the location of single excitation at initial time. Thus the combination of bound state and disorder provides a ideal mechanism for quantum memory.

Although our discussion focuses on single excitation and the general and exact calculations is up to $N=12$, some interesting observation can be found. The main difficulty in the generalization into multi-excitation and large particle number, is to decide the bound state effectively in many-body systems  since the exact solution to Eq. \eqref{boundeqn} become difficult. It is left to address this  interesting issues  in future works. More interestingly  with respect of recent MBL experiments in two dimensional systems \cite{mbl2d}, it is more meaningful whether the similar phenomena can occur since  two dimensional system is more closed to the realization in experiments.

\begin{acknowledgments}
The author HTC acknowledges the support of NSF of China (Grant No. 11005002) and and NCET of Ministry of Education of China (Grant No. NCET-11-0937). HZS acknowledges the support of National Natural Science Foundation of China (NSFC) under Grants No.11705025, China Postdoctoral Science Foundation under Grant No.2016M600223 and No.2017T100192, and the Fundamental Research Funds for the Central Universities under Grant No.2412017QD005. XXY acknowledges the support of NSFC under Grant No.11534002,No.61475033,and No.11775048.
\end{acknowledgments}

\end{document}